# Imaging Inter-Edge State Scattering Centers in the Quantum Hall Regime

Michael T. Woodside,[1] Chris Vale,[1] Paul L. McEuen,[1] C. Kadow,[2] K. D. Maranowski,[2] A. C. Gossard[2]

[1]*Department of Physics, University of California and Materials Science Division, Lawrence Berkeley National Laboratory, Berkeley, California 94720*

[2]*Materials Department, University of California, Santa Barbara, California 93106*

(February 24, 2000)

We use an atomic force microscope tip as a local gate to study the scattering between edge channels in a 2D electron gas in the quantum Hall regime. The scattering is dominated by individual, microscopic scattering centers, which we directly image here for the first time. The tip voltage dependence of the scattering indicates that tunneling occurs through weak links and localized states.

After years of study and two Nobel prizes, the quantum Hall effect continues to provide important challenges to both experimentalists and theorists. Many of the most interesting questions concern the non-uniform spatial structures that can occur within a two dimensional electron gas (2DEG) in high magnetic fields. These structures arise from competition between the effects of Landau level (LL) quantization, Coulomb interactions, and external potentials and include striped phases [1] and insulating phases in the bulk [2] as well as conducting states localized at the edges of the sample (edge states) [3,4]. Scanned probe techniques offer a new approach to investigate these structures directly. They have recently been used to probe the Hall voltage profile and the properties of the insulating state within a quantum Hall plateau [5-8]. Here we use a scanned probe to investigate the microscopic effects of the spatial structure in a 2DEG on electron transport by examining the nature of the scattering between edge states in a quantum Hall conductor.

Transport in the integer quantum Hall regime is now well understood in terms of transport through both quasi-1D edge channels [9] and the bulk. The edge channels are extended states that form along the sample edge due to the effect of the confinement potential on the LL energies. Current can be selectively injected into these edge states [10], creating non-equilibrium edge state (NES) populations that can persist over long distances [11]. Several studies have investigated the length-scales over which NES populations equilibrate [12], and indirect evidence for individual scattering events has been found [13,14]. It has proven difficult, however, to investigate directly the microscopic properties of the scattering. Basic questions remain about the nature of the scattering sites, the frequency with which they occur, and the amount of scattering at each site. Using an atomic force microscope (AFM) tip as a gate to influence inter-edge state scattering, we address these issues by imaging and characterizing individual scattering centers, to our knowledge for the first time. We find that scattering involves tunneling through both through weak links and localized states. These measurements yield a clearer picture of the nature of edge state scattering and also provide lessons about how a scanned probe tip influences a sample.

The sample we study is a GaAs/AlGaAs heterostructure grown by molecular beam epitaxy with a 2DEG lying 90 nm below the surface [15]. The device is patterned into 10 µm-wide Hall bars by wet chemical etching of the heterostructure. The 2DEG has a density of $2.4 \times 10^{15}$ m$^{-2}$ and a mobility of 19 m$^2$/Vs. The device is characterized by standard transport measurements. All measurements are made at temperatures between 0.7 and 1 K and at filling factors between $\nu=3$ and $\nu=2.5$. At these filling factors there are 2 spin resolved outer edge states and a single inner edge/bulk state, as shown in Fig. 1. These filling factors are known to allow significant NES populations [11].

We use two different methods to establish and detect NES populations. In the first method (Fig. 1(a)), a metal gate on top of the 2DEG (the injector gate) is used to selectively inject a non-equilibrium current distribution into the outer edge states [10]. A second gate (the detector gate) is then used to detect the existence of the NES population by selectively transmitting the outer edge states to a voltage probe. The second technique (Fig. 1(b)) uses the fact that NES populations arise naturally in the transition regions between quantum Hall plateaus when the edge and bulk states are decoupled [11]. The non-equilibrium edge states carry excess current that depresses the longitudinal resistance $R_{xx}$. Additional equilibra-tion between the bulk and the edges leads to an increase in $R_{xx}$. Changes in the NES population can thus be detected by measuring $R_{xx}$.

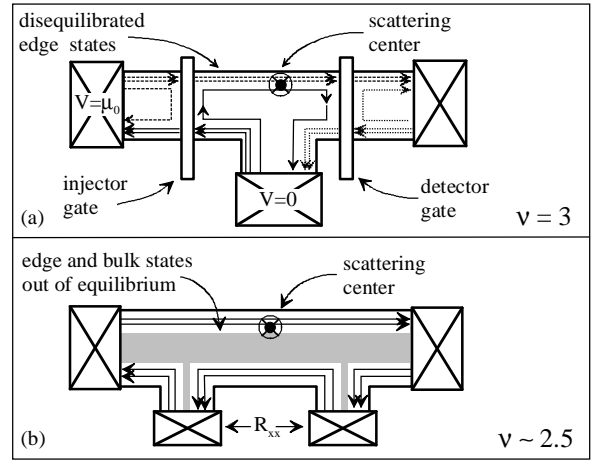

FIG.1. Measurement configurations. (a) Injector gate creates NES population by reflecting $\nu=3$ edge state. Detector gate transmits only $\nu=1,2$ edge states to voltage probe. Equilibration rate determined by comparing probe potential when detector gate reflects $\nu=3$ to probe potential when gate transmits $\nu=3$ (Ref. 10). (b) At $\nu\sim2.5$, where bulk and edge states decouple, excess current in edge states due to NES population suppresses $R_{xx}$. Scattering from edge to bulk increases $R_{xx}$.

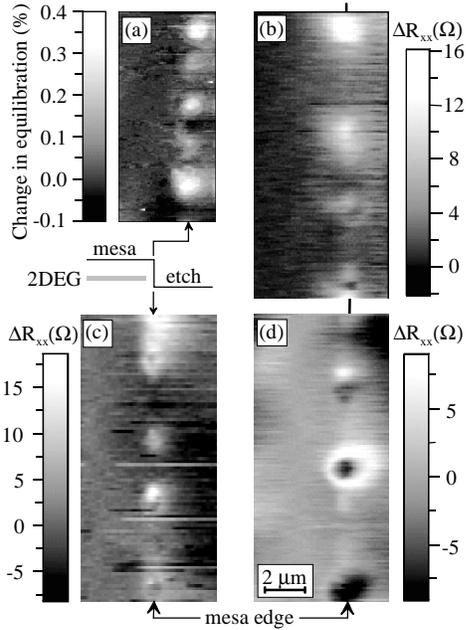

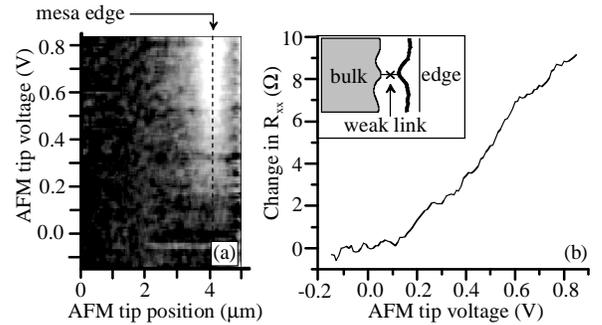

FIG. 3. (a) Cross-section through spot of increased scattering as function of $V_{tip}$. (b) Monotonic increase of scattering at center of feature with $V_{tip}$ suggests tunneling across weak link in incompressible strip (inset).

FIG. 2. (a) Measurement of equilibration between edge states along 10 μm section of 2DEG using the method of Fig. 1(a). Image made with 0.2 mV rms on injector contact and tip bias $V_{tip}$=0.9 V. Bright regions show where AFM tip increases inter-edge state scattering. Features occur at etched edge of mesa, as shown in inset. (b)-(d) Measurement of equilibration between edge and bulk along 3 different 15 μm sections of 2DEG at ν=2.6-2.7 using the method of Fig. 1(b). Bright spots of increased scattering, dark spots of decreased scattering, and bright rings of increased scattering surrounding regions of decreased scattering observed along sample edge. Images made with 100 μA rms current and $V_{tip}$=0.8 V. Scale bar is the same on all images.

Having established a NES population by one of these methods, we use an AFM mounted on a $^3$He cryostat [16] to study the local scattering in the sample. The AFM tip is metallized with Ti and has a diameter of approximately 100 nm. When the tip is biased with a voltage $V_{tip}$, it acts as a local gate [17] and perturbs the states at the edge of the 2DEG. This alters the scattering between edge channels, changing the equilibration rate. We scan the tip 50-75 nm above the sample and measure the equilibration with one of the methods described above. Due to the choice of a low-mobility sample, there is already significant equilibration even in the absence of the AFM tip. What these scanned gate experiments probe is the change in equilibration induced by the tip.

Fig. 2(a) displays the results for a 10 μm long section of the edge of the Hall bar where the NES population is established and detected using top gates. The tip voltage is $V_{tip}$ = 0.9 V [18]. Regions where the scattering is enhanced by the presence of the tip are light, whereas regions where it is reduced are dark. Several bright features representing areas of increased scattering are visible along the edge of the sample. They are not correlated with any topographic features, and they are not observed when the edge and bulk are in equilibrium. They are clearly associated with individual scattering centers, separated on average by ~2 μm.

Similar results are seen when the NES population is established by selective back-scattering of the bulk state, as in Fig. 1(b). Since these measurements do not have to be made between the gates, larger areas can be explored. Figs. 2(b)-(d) show the scattering-induced change in $R_{xx}$ over three different 15 μm long segments of the sample edge at filling factor ν ~ 2.6-2.7. In addition to bright spots of increased scattering, there are dark spots of decreased scattering, and bright rings of increased scattering surrounding regions of decreased scattering. Again, these are observed only along the edge of the sample and are separated on average by a distance of ~2 μm. We find that the bright spots occur most frequently, while the dark spots and the bright rings each occur only about one third as frequently. On average, then, a positive AFM tip bias increases inter-edge state scattering, but at any particular site it can either enhance or reduce the scattering.

The nature of these scattering centers can be probed further by examining the tip voltage dependence of the scattering. A cross-section through the center of one of the bright spots as a function of $V_{tip}$ is shown in Fig. 3(a). As $V_{tip}$ is reduced from positive values, the amount of scattering at the center of the spot decreases monotonically until the spot disappears (Fig. 3(b)). The width at half-maximum of the spot remains roughly constant as $V_{tip}$ is changed. In some cases, a dark spot appears at negative $V_{tip}$, indicating reduced scattering.

Figs. 4 and 5 show the tip voltage dependence of one of the rings of scattering. A cross-section through the ring as a function of $V_{tip}$ (Fig. 4(a)) shows strikingly different behavior from that seen in Fig. 3. There is a strong peak in the amount of scattering at the center of the ring as $V_{tip}$ is changed, as seen in Fig. 4 (b). The evolution of the scattering with $V_{tip}$ revealed by Fig. 4(a) is more clearly illustrated in Fig. 5 by a series of images of the same scattering feature at different tip voltages. As $V_{tip}$ is reduced from positive values, the radius of the ring shrinks linearly with $V_{tip}$ until the ring collapses into a spot. The magnitude of the scattering peak remains constant during this process. As $V_{tip}$ is reduced further, the spot of increased scattering first vanishes and then is replaced by a spot of decreased scattering at negative $V_{tip}$.

These results can be understood in terms of two different scattering mechanisms. Equilibration of the NES populations involves tunneling across the ν=2 incompressible strip between edge and bulk states. The tunneling rate is proportional to $\exp(-a^2/l_B^2)$, where $a$ is the width of the strip and $l_B$ is the magnetic length [19]. Because $a \gg l_B$, tunneling in the quantum Hall regime is normally strongly suppressed. Previous work has shown that the scattering rate can be changed by using a gate alongside the 2DEG to change the confining potential and alter the width of the incompressible strip [4,20]. Positive side gate bias decreases the width of the strip, increasing the equilibration rate, while negative bias has the opposite effect.

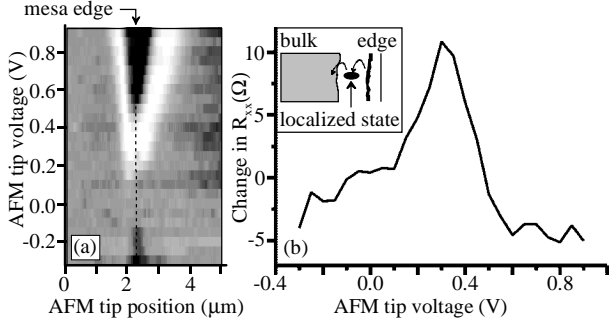

FIG. 4. (a) Cross-section through ring of increased scattering as function of $V_{tip}$. (b) Peak in scattering at center of ring as function of $V_{tip}$ suggests tunneling via localized state (inset).

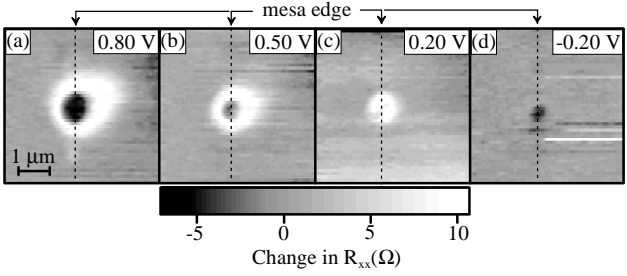

FIG. 5. Ring of increased scattering from Fig. 4 at different $V_{tip}$: (a) 0.8 V, (b) 0.5 V, (c) 0.2 V, (d) –0.2 V. As $V_{tip}$ decreases from positive values, the ring first shrinks, then collapses to a spot. Scattering is reduced by tip at negative $V_{tip}$.

The bright spots of increased scattering (as in Fig. 3) can be understood in terms of tunneling through weak links in the incompressible strip. The width of the strip fluctuates along the length of the sample due to local variations in the electron density of the 2DEG and the roughness of the etched mesa edge, as shown in the inset to Fig. 3(b). Locations where the strip is especially narrow give rise to weak links across which tunneling occurs preferentially. As with experiments using side gates, the AFM tip bias perturbs the width of the incompressible strip. A positive tip bias hardens the confining potential and decreases the width of the strip, increasing tunneling through the weak link. The scanned gate results are thus consistent with previous work showing an increase in equilibration rates with more positive side gate bias. The high spatial resolution of the AFM tip, however, reveals that the increased scattering occurs only at specific sites along the edge.

The very different behavior seen in Fig. 4 indicates that a different scattering mechanism is involved. The scattering rings can be understood in terms of tunneling through a quasi-bound localized state in the incompressible strip (inset to Fig. 4(b)). This behavior is similar to the resonant scattering phenomena observed in narrow quantum Hall conductors, where electrons scatter from one edge to the other via localized states in the bulk [21]. The localized states arise possibly from the donor impurity potential or from defects near the heterostructure interface. The AFM tip electrostatically alters the energy of the localized state. At some value of $V_{tip}$ the localized state is brought into resonance with the Fermi level $E_F$, causing a peak in the scattering as seen in Fig. 4(b). The annular scattering patterns (Fig. 5) thus map out the locations at which the tip brings the localized state into resonance with $E_F$ at any given tip bias. Since the amount of scattering depends only on how close the level energy is to $E_F$, the height of the scattering peak is independent of $V_{tip}$.

We see from Fig. 4(b) that the unperturbed level lies just above $E_F$. Negative $V_{tip}$ moves the level further away from $E_F$, decreasing the scattering (Fig. 5(d)). A small positive $V_{tip}$ aligns the level with $E_F$, causing a single scattering peak over the location of the scatterer (Fig. 5(c)). As $V_{tip}$ increases, the equipotential contour at $E_F$ for the localized level expands out from the scatterer, changing the single peak into a ring whose radius grows with $V_{tip}$ (Figs. 5(a-b)) [22].

We now turn to the question of how much equilibration occurs at each site. This can be determined from the results of Fig. 2(a), where top gates are used to measure the non-equilibrium potential difference $\Delta \mu$ between the ν=1,2 and ν=3 edge states. In the absence of the tip, the total equilibration rate over the 30 μm distance between top gates is ~90%. Since scattering sites are located every ~2 μm, the average scattering probability $p$ needed to account for the measured equilibration rate [14] is $p \sim 0.15$. The NES population is therefore reduced by ~15% at each microscopic scattering site.

We can also determine from Fig. 2(a) the amount of extra scattering caused at each scattering site by the AFM tip perturbation. The change in the scattering probability $\Delta p$ induced at a particular site is given by the fractional change in $\Delta \mu$ caused by the tip at that site. For the scattering sites observed in Fig. 2(a) with $V_{tip} = 0.9$ V, we find that $\Delta p = 0.1-0.3$, with an average value of $\Delta p = 0.2$. The amount of scattering induced by the tip at this tip bias is thus of the same order as the scattering already present in the sample.

These experiments are, to our knowledge, the first direct measurement of the amount of edge state coupling at individual scattering sites. They show that the equilibration is dominated by strong scattering centers separated by a few μm. This contrasts with the results of a previous study, which inferred the existence of scattering sites with $p \sim 0.006-0.02$ separated by ~90-600 nm based on a statistical analysis of scattering between the ν=2 and ν=1 (spin-polarized) edge states [14]. The origin of the differences between these two experiments is not clear. However, the momentum and spin conservation issues for scattering between edge states of different orbital LLs are very different from those for scattering between different spin states within the same LL [9].

The measurements presented here clearly probe individual scattering centers. We emphasize, however, that the relation between the observed features and the underlying scattering centers is not straightforward, due to the complex tip-sample electrostatics. As discussed above, the annular structures observed in Fig. 5 clearly do not correspond to annular scattering centers, but rather to equipotential contours around a single scattering center. The ring of scattering in Fig. 5 is also not circular, as might naively be expected, but flattened on the side over the mesa. This is due to the spatial variation in the dielectric properties of the sample near the mesa edge. The high-dielectric GaAs ($\varepsilon \sim 13$) and the 2DEG screen the tip more effectively when the tip is over the mesa than when it is over the etched region, flattening the side of the ring over the mesa. We note as well that the scanned gate features in Figs. 2-5 are very near the physical edge of the sample. Previous theoretical [4] and experimental work [23], however, indicates that the edge states reside several hundred nm inside the mesa due to depletion of the 2DEG near the sample edge. We again attribute this to the non-uniform screening properties near the sample edge: the tip has its greatest effect near the edge of the mesa where the 2DEG and the GaAs are less effective in screening it. This further illustrates the complexities in relating features observed in scanned probe images to the underlying spatial structures in the 2DEG.

In summary, we have for the first time imaged individual scattering centers responsible for the equilibration of edge state populations in a quantum Hall conductor. By studying the dependence of the scattering on tip voltage, we find evidence for tunneling across weak links and tunneling through localized states. This work clearly illustrates that scanned gate microscopy can be a powerful tool for probing inhomogeneous microscopic structures in quantum Hall conductors.

This work was supported by the NSF, NSERC, the AT&T Foundation, and the Packard Foundation. We acknowledge the Berkeley Microlab for sample fabrication.

———————————

[1] M. P. Lilly *et al.*, Phys. Rev. Lett. **82**, 394 (1999).
[2] For a review, see S. Sondhi *et al.*, Rev. Mod. Phys. **69**, 315 (1997).
[3] B. I. Halperin, Phys. Rev. B **25**, 2185 (1982); M. Buttiker, Phys. Rev. B **38**, 9375 (1988).
[4] D. B. Chklovskii, B. I. Shlovskii, and L. I. Glazman, Phys. Rev. B **46**, 4026 (1992).
[5] K. L. McCormick *et al.*, Phys. Rev. B **59**, 4654 (1999).
[6] A. Yacoby *et al.*, Solid State Comm. **111**, 1 (1999).
[7] S. H. Tessmer *et al.*, Nature **392**, 51 (1998).
[8] G. Finkelstein *et al.*, cond-mat 9910061.
[9] For a review, see R. J. Haug, Semicond. Sci. Tech. **8**, 131 (1993).
[10] B. J. van Wees *et al.*, Phys. Rev. Lett. **62**, 1181 (1989).
[11] B. J. van Wees *et al.*, Phys. Rev. B **39**, 8066 (1989); B. W. Alphenaar *et al.*, Phys. Rev. Lett. **64**, 677 (1990).
[12] S. Komiyama *et al.*, Phys. Rev. B **45**, 11085 (1992); G. Müller *et al.*, Phys. Rev. B **45**, 3932 (1992).
[13] B. W. Alphenaar, P. L. McEuen, and R. G. Wheeler, Physica B **175**, 235 (1991).
[14] Y. Acremann *et al.*, Phys. Rev. B **59**, 2116 (1999).
[15] The heterostructure is the same as sample A in K. D. Maranowski *et al.*, Appl. Phys. Lett. **66**, 3459 (1995).
[16] K. L. McCormick, Ph. D. thesis, University of California, Berkeley, 1998.
[17] M. A. Eriksson *et al.*, Appl. Phys. Lett. **69**, 671 (1996).
[18] Large $V_{tip}$ (outside the range ±1 V) causes abrupt, hysteretic changes in the scattering. This is likely due to rearrangement of charges in the donor or surface layers (Refs. 7,8).
[19] T. Martin and S. Feng, Phys. Rev. Lett. **64**, 1971 (1990).
[20] R. J. F. van Haren *et al.*, Phys. Rev. B **47**, 15700 (1993).
[21] J. K. Jain and S. A. Kivelson, Phys. Rev. Lett. **60**, 1542 (1988); P. C. Main *et al.*, Phys. Rev. B **50**, 4450 (1994); D. H. Cobden, C. H. W. Barnes, and C. J. B. Ford, Phys. Rev. Lett. **82**, 4695 (1999).
[22] Localized levels below $E_F$ when unperturbed should display the same behavior as in Fig. 5 but at the opposite polarity of $V_{tip}$. We believe this is the origin of the dark spots of reduced scattering at positive $V_{tip}$ in Fig. 2. It also explains why such dark spots occur with the same frequency as the rings of increased scattering.
[23] Y. Y. Wei *et al.*, Phys. Rev. Lett. **81**, 1674 (1998).